\documentclass[reprint,amsmath,amssymb,aps, twocolumn]{revtex4-2}

\usepackage{hyperref}
\usepackage{comment}
\hypersetup{colorlinks,urlcolor=blue}
\usepackage{graphicx}
\usepackage[percent]{overpic}
\usepackage{dcolumn}
\usepackage{bm}
\usepackage{ulem}
\usepackage{physics}
\usepackage{lipsum}
\usepackage{appendix}
\usepackage{overpic}
\usepackage{enumerate}
\usepackage[T1]{fontenc}

\newcommand{\ham}{\mathcal{H}}
\newcommand{\vect}[1]{\boldsymbol{#1}}

\begin{document}
\preprint{APS/123-QED}
\title{
Spin-spiral instability of the Nagaoka ferromagnet in the crossover between square and triangular lattices
}
\author{Darren Pereira}
\email{dlp263@cornell.edu}
\author{Erich J. Mueller}
\email{em256@cornell.edu}
\affiliation{Laboratory of Atomic and Solid State Physics, Cornell University, Ithaca, New York 14853, USA}

\begin{abstract}
We study the hard-core Fermi-Hubbard model in the crossover between square and triangular lattices near half-filling.  As was recognized by Nagaoka in the 1960s, on the square lattice the presence of a single hole leads to ferromagnetic spin ordering.  
On the triangular lattice, geometric frustration instead leads to a spin-singlet ground state, which can be associated with a 120-degree spiral order.  On lattices which interpolate between square and triangular, there is a phase transition at which the ferromagnetic order becomes unstable to a spin spiral.  We model this transition, finding the exact location of the spin-spiral instability.
\end{abstract}

\date{\today}
\maketitle

\section{Introduction and Overview}
\label{sec:Intro}

Placing a single hard-core fermion on each site of a lattice leads to a highly-degenerate jammed insulating state, where all spin configurations have the same energy.  
Adding a single mobile hole allows the spins to rearrange themselves, breaking this degeneracy.  
In a landmark paper, Nagaoka proved that on a class of lattices that includes the two-dimensional square lattice the resulting ground state is ferromagnetic, with maximal total spin \cite{NagaokaPR1966}; a similar finding was also made by Thouless \cite{ThoulessProc1965}.
For spin-1/2 particles on a triangular lattice the ground state is instead a spin singlet \cite{HaerterPRL2005}, which can be interpreted as a spiral pattern where the spins on each of the three sublattices are rotated by 120$^\circ$ with respect to one another. A natural question is how these two spin orders are connected as one changes the lattice geometry from square to triangular. 
Prior numerical studies suggested that a phase transition occurs as one deforms the square lattice geometry, marked by the formation of a long-wavelength spin spiral \cite{LisandriniPRB2017,SharmaArxiv2025,PereiraArxiv2025}.
Here, we develop an analytic model of this transition, showing that the leading instability of the uniform ferromagnet is indeed towards a spin spiral. Under the assumption that no other instabilities preempt this transition, we determine the exact critical point.

To interpolate between lattice geometries we consider a hopping Hamiltonian $\ham=-\sum_{ ij,\sigma}  t_{ij} c_{i\sigma}^\dagger c_{j\sigma}$. Here $c_{j\sigma}$ is the annihilation operator for a fermion of spin $\sigma$ on site $j$, $t_{ij}=t_{ji}$, and an implicit hard-core constraint forbids two fermions from occupying the same site. 
As depicted in the inset of Fig.~\ref{fig:Phases}, we  arrange our sites on a square lattice, taking $t_{ij}=t$ for neighboring fermions in the cardinal directions and $t_{ij}=t^\prime$ for neighbors along one diagonal.  The case $t^\prime=0$ corresponds to the standard square lattice, while $t^\prime=t$ is equivalent to a triangular lattice.  Up to a redefinition of our momentum vectors, we would find identical results if we instead arranged our sites on a triangular lattice, as in Refs.~\cite{HaerterPRL2005, LisandriniPRB2017, SharmaArxiv2025}.

The key physics of the transition is then captured at the mean-field level by a simple
variational ansatz, 
\begin{align}\label{psi0}
|\psi\rangle=\sum_i f_i \prod_{j\neq i} (u_j  c_{j\uparrow}^\dagger +  v_j  c_{j\downarrow}^\dagger) |{\rm vac}\rangle,
\end{align}
describing the motion of a single hole in the presence of a static spin pattern.
The spin direction on site $j$ is encoded by $(u_j,v_j)=[\cos(\theta_j/2),\sin(\theta_j/2)]$, and $f_i$ corresponds to the amplitude for the hole to be on site $i$.  
For sufficiently small $t^\prime$ the variational energy is minimized by a uniform spin pattern, with $f_j=e^{i \vect{k}\cdot \vect{r}_j}$ for $\vect{k}=(\pi,\pi)$, corresponding to putting the hole at the top of the band.  For $t^\prime>(t_{\rm c}^\prime)_{\rm MF}=t/2$ this variational ansatz has a large degenerate ground-state manifold, representing a range of different possible spin patterns. Spin fluctuations break this degeneracy. We calculate the leading contribution from these fluctuations, finding that they favor spin-spiral patterns, $\theta_j=\vect{ Q}\cdot \vect{r}_j$.  They also shift the transition point to $(t_{\rm c}^\prime)_{\rm exact}=0.24 t$. Figure ~\ref{fig:Phases} summarizes the findings of this Letter. 

In the ferromagnetic phase the exact ground state is very simple, corresponding to a single hole in a spin-polarized background. The picture is more complicated in the spin-spiral phase. A spin texture provides a Berry phase for the hole's motion \cite{SposettiPRL2014}, which is partly responsible for the mean-field degeneracy.  For states of the form of Eq.~(\ref{psi0}), the momentum of the hole can be shifted by instead twisting the spin configuration. 
When fluctuations are added, the spiral patterns have the lowest energy because they most effectively couple the hole's motion with the spin excitations.

Nevertheless, this complexity is irrelevant to finding the critical point.  {One can find the phase boundary by simply looking at the stability of the ferromagnetic state} against forming a long-wavelength spin spiral (i.e. by searching for a vanishing spin stiffness).    In a recent paper, Sharma \textit{et al.} made a similar observation \cite{SharmaArxiv2025}, 
estimating the location of the phase transition by looking at the spin-wave instability of the ferromagnet. 
As they observed, this spin-wave argument gives an incorrect value for the critical point, $(t_{\rm c}^\prime)_{\rm SW}=0.42 t$, significantly larger than the critical hopping strength that they found in their numerical calculations.
This overestimation occurs because the actual instability is towards a spiral, rather than a spin wave.  Note that they use a different orientation of their lattice, so some of their expressions look slightly different than ours.

Although both spin waves and spin spirals correspond to slow 
spatial rotations of the spins, they are very different excitations. In  a spin wave the spins oscillate about a fixed direction, for example staying near one of the poles of the Bloch sphere.  Conversely, in a spiral the spins continuously rotate with a fixed pitch, tracing out a great circle.  Quantum mechanically, a ferromagnet of $N$ particles with a single spin wave has a total spin $S=N/2-1$, and can be constructed by flipping a single spin.  Conversely, a spin spiral has $S=0$, and requires flipping a macroscopic number of spins.  
  
This phase transition was first studied numerically by Lisandrini \textit{et al.} \cite{LisandriniPRB2017}, using density-matrix renormalization group techniques on finite lattices.  
Later Sharma \textit{et al.} \cite{SharmaArxiv2025} extended these calculations to larger system sizes, and generalized previous analytic approaches \cite{DavydovaPRB2023}.  
At finite temperature this phase transition becomes a crossover, as was studied by the current authors \cite{PereiraArxiv2025}.
A variant of our 
present
approach has 
been used to calculate the superfluid drag in a model of two-component bosons \cite{KielyPRA2025}.

It is important to note that we are working in the hard-core limit, where two particles cannot occupy the same site.  This constraint could be relaxed, replacing it with an on-site Hubbard repulsion of strength $U$.  In that setting, superexchange with scale $J\sim t^2/U$ will compete with the effects of kinetic magnetism, leading to polaronic physics where the spin ordering depends on the distance from the hole \cite{WhitePRB2001, DavydovaPRB2023}.   In this Letter we avoid these complications by setting $U\to\infty$.

Cold-atom experimentalists have demonstrated tunable optical lattices which
interpolate between square and triangular geometries in exactly the way we consider here \cite{XuNature2023}. 
Using site-resolved imaging, they have explored magnetic correlations in the vicinity of individual holes \cite{LebratNature2024, PrichardNature2024}, but not in the crossover between geometries.  Although superexchange was important in those experiments, it is possible to extend the experiments further into the strongly-interacting regime to directly explore the kinetic magnetism effects discussed here.  
This physics may also be explored in other settings, such as coupled arrays of transmons 
\cite{Kjaergaard2020} or moir\'e superlattices formed from layering exfoliated two-dimensional materials \cite{WuPRL2018, CiorciaroNature2023, TaoNatPhys2024}.  A broader review of the literature can be found in Ref.~\cite{PereiraArxiv2025}. 

\section{Calculation} 
\label{sec:Setup}

\begin{figure}[t!]
    \centering        
    \includegraphics[width=\linewidth]{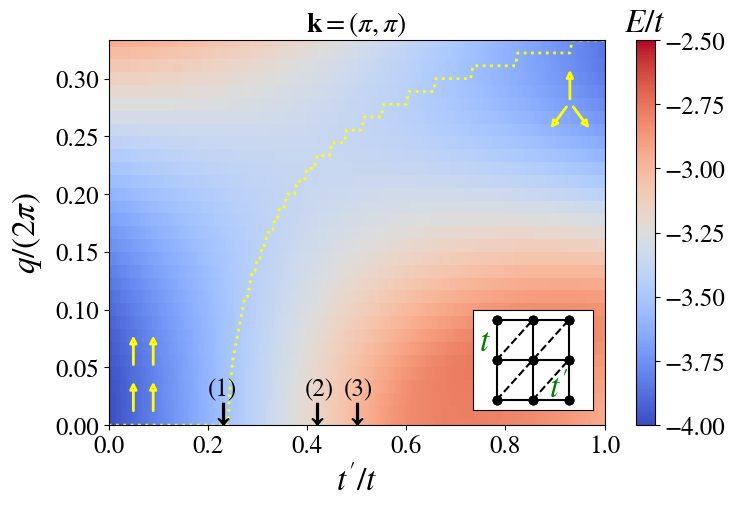}
    \caption{
    Spin-spiral instability of the ground state of the half-filled hard-core Fermi-Hubbard model, with a single hole.  Inset shows hopping model where $t$ corresponds to hopping matrix element along cardinal directions, and $t^\prime$ along diagonals with positive slope.  This interpolates between a square lattice at $t^\prime=0$ and a triangular lattice at $t^\prime=t$.  Main figure shows the variational energy of a spin-spiral configuration, as calculated from Eqs.~\eqref{eq:Helements1} and \eqref{eq:Helements2} on a small 
    $15\times 15$ spatial grid.  Vertical axis shows $q$ where the  spiral has wave-vector $\vect{Q} = (q,q)$. Horizontal axis corresponds to the hopping ratio, $t^\prime/t$.  For each $t^\prime/t$ a yellow dot is placed at the $q$ which minimizes the energy.  The variational calculation is uncontrolled at finite $q$, but becomes exact as $q\to 0$, and hence it predicts the exact location of the spin-spiral instability, labeled by (1):
    $(t_{\rm c}^\prime)_{\rm exact} = 0.24t$.
    This can be contrasted to the spin-wave instability (2), occurring at $(t_{\rm c}^\prime)_{\rm SW} = 0.42t$, and the mean-field prediction (3), $(t_{\rm c}^\prime)_{\rm MF} = 0.5t$.  In the lower left and upper right corner, one unit cell of the $t^\prime=0$ and $t^\prime=t$ spin patterns are depicted, corresponding to the uniform ferromagnet and the 120-degree state.
    }
    \label{fig:Phases}
\end{figure}

Our goal is to find the parameter $t^\prime/t$ at which the uniform ferromagnet becomes unstable to forming a spin spiral with wave-vector $\vect{Q}=(q,q)$.  
Since $q=0$ at the transition, it suffices to calculate the energy, $E(t,t^\prime,\vect{Q})=E_0(t,t^\prime)+q^2 E_2(t,t^\prime)+\cdots$, to quadratic order in $q$.  The transition occurs when the spin stiffness vanishes, $E_2(t,t^\prime)=0$.  In the Supplemental Material \cite{supp} we consider more general spin patterns and demonstrate that they are all of higher energy. We also present an argument that the wave-vector has this form.
While it does not affect the physics, the definition of the wave-vector depends on how the lattice sites are arranged in space.  Our $\vect{Q}=(q,q)$ on the square lattice becomes $\vect{Q}_{\rm t}=(q,0)$ on the triangular lattice, which is the geometry used in Refs.~\cite{HaerterPRL2005, LisandriniPRB2017, SharmaArxiv2025}.

We begin with the ansatz in Eq.~\eqref{psi0}.  The spin-spiral solutions correspond to $f_j=e^{i\vect{k}\cdot \vect{r}_j}$ and $\theta_j={\vect{Q}\cdot \vect{r}_j}$.  
When $q=0$, this is an eigenstate of the Hamiltonian.  Thus for small $q$, it is nearly an eigenstate, and we can use perturbation theory to {\it exactly} calculate $E_2$ and the critical point.

It is convenient to
make a local basis change,
where the spin on every site is rotated into the local frame of reference of the spin spiral:
\begin{equation}
  \begin{bmatrix}
    c_{i\uparrow} \\ c_{i\downarrow}
  \end{bmatrix} = 
  \begin{bmatrix}
    \cos(\vect{Q} \cdot \vect{r}_i/2) & -\sin(\vect{Q} \cdot \vect{r}_i/2) \\
    \sin(\vect{Q} \cdot \vect{r}_i/2) & \cos(\vect{Q} \cdot \vect{r}_i/2)
  \end{bmatrix} 
  \begin{bmatrix}
    b_{i} \\ a_{i}
  \end{bmatrix}.
\end{equation}
We then perform a particle-hole transformation on  the $b$-fermions, $h_{i} = b_{i}^\dagger$ and $h_{i}^\dagger = b_{i}$, while keeping the $a$ operators untouched. Our transformed Hamiltonian becomes ${\cal H}=
{\cal H}_0+{\cal H}^\prime$ with
\begin{align}\label{h0}
    {\cal H}_0 &= -\sum_{ ij } t_{ij} \cos(\vect{Q} \cdot \vect{r}_{ij}/2) (-h_{i}^\dagger h_{j} + a_{i}^\dagger a_{j} ), \\
    \label{hp}
    {\cal H}^\prime&=- \sum_{ij} t_{ij} \sin(\vect{Q} \cdot \vect{r}_{ij}/2) (h_{i}a_{j} - a_{i}^\dagger h_{j}^\dagger ).
\end{align}
Although this is a quadratic Hamiltonian, it is not exactly solvable, as we also have a hard-core constraint.  Nonetheless, we can take advantage of the fact that ${\cal H}^\prime$ vanishes as $q\to 0$, giving us a controlled expansion.  The 
leading contribution from ${\cal H}^\prime$ is captured by
the variational ansatz,
\begin{equation}
  \ket{\psi_1}=\frac{1}{\sqrt{N_s}}\sum_l
  e^{i \vect{k} \cdot \vect{r}_l}
  \left(f_0+\sum_{s\neq 0}f_s  a_{l+s}^\dagger h_{l+s}^\dagger \right) h_l^\dagger|{\rm vac}_q \rangle,\label{psivar}
\end{equation}
 where $|{\rm vac}_q\rangle$ is the vacuum state annihilated by $h_i$ and $a_i$, corresponding to a pristine spin spiral.
The coefficient $f_0$ represents the amplitude to have  a single hole, while $f_s$ 
represent amplitudes for  also having a flipped spin (in the rotated basis) a distance $s$ from the hole.  To zeroth order in $\ham^\prime$ these coefficients vanish, and $f_s\propto q$ as $q\to 0$.  Higher-order contributions from ${\cal H}^\prime$ would result in more flipped spins and require a more sophisticated wavefunction.  The spin-wave ansatz in Refs.~\cite{SharmaArxiv2025, DavydovaPRB2023} is a special case of Eq.~(\ref{psivar}), with $f_0=0$ and $\vect{k}=\vect{Q}=\vect{0}$.

The energy minimum  occurs at $\vect{k}=(\pi,\pi)$ and $\vect{Q}=(q,q)$, but our argument applies for generic wave-vectors.
Minimizing the variational energy yields a set of equations
\begin{align}
    E f_0&= \sum_{\vect{u}} C_{\vect{u}} f_0 - f_{\vect{u}} S_{\vect{u}}, 
    \label{eq:Helements1}
    \\\label{eq:Helements2} 
    Ef_{\vect{s}} &= \sum_{\vect{u}} \left(S_{\vect{u}}f_0 + f_{\vect{u}} C_{\vect{u}}\right)\delta_{\vect{s}, -\vect{u}} + f_{\vect{u} + \vect{s}} (1-\delta_{\vect{s}, -\vect{u}}) C_{\vect{u}}. 
\end{align}
Here, and in future expressions, $\vect{u}$ runs over the six nearest neighbors of the origin: $\vect{u}\in\{$$\pm \vect{x}, \pm \vect{y},$ $\pm \vect{w}\}$ where $\vect{x}$ and $\vect{y}$ are unit vectors, and  $\vect{w} \equiv \vect{x} + \vect{y}$.  We have also defined
\begin{align}\label{s}
    S_{\vect{u}} &= e^{-i\vect{k} \cdot \vect{u}} t_{\vect{u}} \sin(\vect{Q}\cdot \vect{u}/2),\\
    C_{\vect{u}} &= e^{-i\vect{k} \cdot \vect{u}} t_{\vect{u}} \cos(\vect{Q} \cdot \vect{u}/2),
\end{align}
where $t_{\pm \vect{x}}=t_{\pm \vect{y}}=t$ and $t_{\pm \vect{w}}=t^\prime$.
One can interpret Eq.~(\ref{eq:Helements1}) 
and (\ref{eq:Helements2}) 
as a single-particle hopping Hamiltonian with a peculiar defect at the origin.  If one truncates to a finite number of sites this eigenproblem can be easily solved numerically, yielding an approximation to $E(t,t^\prime,\vect{Q})$.  Figure~\ref{fig:Phases} shows the resulting energy landscape from this calculation on a small spatial grid. The critical point is clearly visible, where the energy minimum shifts away from $q=0$. 

To analytically find the phase transition, we perturbatively solve Eq.~(\ref{eq:Helements1}) 
and (\ref{eq:Helements2}).  
To zeroth order  in $q$, we have $f_0=1$ and $E=\epsilon_0=\sum_{\vect{u}} C_{\vect{u}}=-4 t+2 t^\prime$.
When ${\vect{Q}}=(q,q)$, the first corrections take the form $E=\epsilon_0
+q^2 (t/2-t^\prime)
-\sum_{\vect{u}} S_{\vect{u}} f_{\vect{u}}$, where $\vect{u}$ runs over the six nearest neighbors.  We find $f_{\vect{u}}$, to linear order in $q$, by substituting the zeroth-order solution into the second equation, and introducing the  
Fourier transforms,
\begin{align}\label{ft}
    f_{\vect{s} \neq 0} &= \frac{1}{\sqrt{N_s}} \sum_{\vect{p}} e^{i \vect{p} \cdot \vect{s}} g_{\vect{p}}, 
    &g_{\vect{p}} &= \frac{1}{\sqrt{N_s}} \sum_{\vect{s} \neq 0} e^{-i\vect{p} \cdot \vect{s}} f_{\vect{s}} .
\end{align}
As discussed in Ref.~\cite{KielyPRA2025} we are free to add an arbitrary constant to $g_p$, as 
the Fourier sum only involves sites where $\vect{s}\neq \vect{0}$.  The number of lattice sites is $N_s$.  We multiply Eq.~(\ref{eq:Helements2}) by $e^{-i{\vect{p}\cdot \vect{s}}}$ and sum over $\vect{s}$ to find an expression for $g_{\vect{p}}$ in terms of the six $f_{\vect{u}}$'s,
\begin{align}
(\epsilon_0-\epsilon_{\vect{p}})g_{\vect{p}}&=\frac{1}{\sqrt{N_s}}\sum_{\vect{u}} e^{i {\vect{ p} \cdot \vect{u}}} \left(S_{\vect{u}}+ f_{\vect{u}} C_{\vect{u}}\right)
\end{align}
where $\epsilon_{\vect{p}}=
\sum_{\vect{u}} t_{\vect{u}} e^{i\vect{p}\cdot\vect{u}} e^{-i\vect{k} \cdot \vect{u}}
$ is the single-particle dispersion.
We then use the inverse Fourier transform from Eq.~(\ref{ft}) to find a closed set of equations relating the $f_{\vect{u}}$'s.  These can be written as
\begin{align}\label{fs}
    \sum_{\vect{u}} S_{\vect{u}} \Lambda_{\vect{s}+\vect{u}}
    &= f_{\vect{s}} + \sum_{\vect{u}} f_{\vect{u}} C_{\vect{u}} \left[\Lambda_{\vect{s}} - \Lambda_{\vect{u} + \vect{s}} \right],
\end{align}
where we have defined
\begin{align}
    \Lambda_{\vect{s}} &= \frac{1}{N_s} \sum_{\vect{p}} \frac{e^{i\vect{p} \cdot \vect{s}}}{\epsilon_0 - \epsilon_{\vect{p}}}.\label{int}
\end{align}
Specializing to the case $\vect{Q}=(q,q)$,
these equations can be simplified by using reflection and inversion symmetries:  $\Lambda_{(s_x,s_y)}=\Lambda_{s_y,s_x}$ and $\Lambda_{ -\vect{s}}=\Lambda_{\vect{ s}}$.  Consequently $f_x=f_y=-f_{-x}=-f_{-y}$ and $f_{-w}=-f_w$, and this reduces  Eq.~\eqref{fs} to two coupled equations.  As shown in the Supplemental Material \cite{supp}, the equations are readily solved: by using the residue theorem, the two-dimensional integrals $\Lambda_{\vect{s}}$ can be converted to one-dimensional integrals, which are easily calculated numerically.  
We find that $E_{2}(t,t^\prime)=0$ when $t^\prime=0.24 t$. 

\section{Summary and Outlook}
\label{sec:Summary}

We studied how a single hole leads to magnetic ordering in a gas of hard-core fermions as one interpolates between a square and triangular lattice, parameterized by the diagonal hopping $t^\prime$.  We began with a simple mean-field ansatz where the hole moves through a static spin texture.  For a square lattice, $t^\prime=0$, the energy is minimized by a ferromagnetic pattern.  This is the exact ground state, as found by Nagaoka \cite{NagaokaPR1966} and Thouless \cite{ThoulessProc1965}.  As one increases $t^\prime$, there is a critical point, beyond which the 
mean-field ground state is highly degenerate.  Fluctuations resolve this degeneracy; we found that the lowest-energy spin pattern is a spin spiral, whose pitch continuously grows from $q=0$ at the transition.  This becomes the
120-degree state in the triangular lattice, $t^\prime=t$, consistent with the observations of Haerter and Shastry \cite{HaerterPRL2005}.  We used second-order perturbation theory to calculate the critical point $t^\prime_{\rm c}=0.24t$.  Given that the mean-field ansatz is exact for the ferromagnet, this perturbative calculation yields the exact critical point, assuming that we have correctly identified the leading instability.

Prior experiments were performed at weaker coupling, where superexchange effects played an important role \cite{XuNature2023, PrichardNature2024, LebratNature2024}.  Nonetheless it is reasonable to envision repeating those experiments with deeper optical lattices, to reach the regime described here.  One could directly image the spin spirals.  For the single-hole problem quantum statistics only enter into the sign of the hopping matrix elements.  Thus one could also envision studying this physics by using bosonic systems, such as arrays of qubits \cite{RevModPhys.86.153}.

Finite-temperature corrections were discussed in our previous work \cite{PereiraArxiv2025}.  There the magnetic ordering is restricted to a region of space near the hole, corresponding to a polaron.  The polaron's size grows as temperature is lowered. There is no magnetic phase transition at these elevated temperatures, but there is a crossover, which can be determined by observing the spin correlations near the hole.  The crossover occurs at a value of $t^\prime$ which is very similar to $t_{\rm c}^\prime$.

Our calculations were performed in the limit of exactly one hole in an infinite system.  Some prior work has explored the case of finite hole density \cite{SharmaArxiv2025}, suggesting  that $t_c^\prime$ should fall with increasing density of holes, $n_h$.  When $n_h<5\%$ they argued that the finite-density corrections should be small.

The problem of kinetic magnetism is over 60 years old.  It is a quintessential example of strongly-correlated physics. Cold-atom quantum simulators allow this and other topics to be studied from a fresh angle, shedding new light on such milestone problems. These engineered systems provide a level of quantum control and measurement that is not available in conventional materials. With such control, it is now possible to systematically explore the interplay between kinetic magnetism and frustration, which is an ideal setting for exotic quantum phenomena \cite{ZhangPRB2018, GlittumNature2025, ZhangArxiv2025}. Our work highlights one consequence of kinetic frustration, a continuous phase transition between a ferromagnet and a spin-spiral texture. It will be exciting for the field to uncover other phenomena that arise from kinetic frustration, and to demonstrate these in experiments.

\section*{Data Availability}
The data which supports this work is available from the authors upon reasonable request.

\section*{Acknowledgements}
We thank Thomas Kiely for constructive comments on the manuscript.
This material is based upon work supported by the National Science Foundation under Grant No.  PHY-2409403. We also acknowledge the support of the Natural Sciences and Engineering Research Council of Canada (NSERC) (Ref. No. PGSD-567963-2022).

%

\clearpage
\section*{Supplemental Material for \\ ``Spin-spiral instability of the Nagaoka ferromagnet in the crossover between square and triangular lattices'', Darren Pereira and Erich J Mueller}

\setcounter{section}{0}
\renewcommand{\theequation}{S\arabic{equation}}
\setcounter{equation}{0}
\renewcommand{\thesection}{S\arabic{section}}
\renewcommand{\thefigure}{S\arabic{figure}}
\setcounter{figure}{0}
\section{Fluctuations about Generic Mean-Field Ansatz}
\label{sec:MFAppendix}

In this Section, we analyze the variational ansatz in Eq.~(\ref{psi0}),
$
|\psi\rangle=\sum_i f_i \prod_{j\neq i} (u_j  c_{j\uparrow}^\dagger +  v_j  c_{j\downarrow}^\dagger) |{\rm vac}\rangle,
$
generalized to complex $u_j$ and $v_j$.
We show that this mean-field ansatz generically yields a family of degenerate solutions.  Quantum fluctuations break this degeneracy.  We argue that  the spin-spiral solutions maximize the energy saved by these fluctuations.  

{\color{black}
In Section~\ref{quad}, we find the form of the mean-field states which extremize the energy.  We calculate their energy, including quadratic fluctuations, as $q\to 0$, and use scaling arguments to show that the leading instability of the ferromagnet is a spin spiral. In Section~\ref{global} we give numerical evidence that this result persists to finite $q$.  In particular, we calculate how the variational energy depends on the spin texture in the triangular-lattice limit, where $q=2\pi/3$.}

\subsection{Calculation of 
Energy to Quadratic Order}
\label{quad}
As detailed in the main text, we consider a hopping Hamiltonian, $\ham=-\sum_{ij\sigma}  t_{ij} c_{i\sigma}^\dagger c_{j\sigma}$ with $t_{ij}=t_{ji}$, and a hard-core constraint, which is automatically satisfied by our variational wavefunction.  The expectation value of the energy is
\begin{align}
    \langle \psi| \ham|\psi\rangle
    &=\sum_{ ij} t_{ij} (\tilde u_j^* \tilde u_i + \tilde v_j^* \tilde v_i),
\end{align}
where $\tilde u_i=f_i^* u_i$ and $\tilde v_i= f_i^* v_i$.  The fact that the energy only depends on these combinations  illustrates a redundancy in our parameterization of the wavefunction.  Minimizing this energy, with the normalization constraint that $\sum_i |\tilde u_i|^2 + |\tilde v_i|^2=1$, results in the eigenproblem
\begin{align}
    \sum_j t_{ij} \tilde u_j&=\epsilon \tilde u_i,
    &\sum_j t_{ij} \tilde v_j&=\epsilon \tilde v_i,
\end{align}
where $\epsilon$ is a Lagrange multiplier, corresponding to the variational energy.  
We take a translationally-invariant hopping $t_{ij}=t_{\vect{r}_i-\vect{r}_j}$.
The solution to the eigenproblem is 
then of the form $\tilde u_i=\tilde u e^{i \vect{p}\cdot \vect{r}_j}$,  $\tilde v_i=\tilde v e^{i {\vect{p} \cdot \vect{r}_j}}$, with energy  $\epsilon=\sum_{\vect{u}} t_{\vect{u}} e^{i {\vect{p} \cdot \vect{r}}_{\vect{u}}}$.  For the special case of square lattice, with hopping $t$ along the cardinal directions and $t^\prime$ only along the diagonals with positive slope, this becomes
\begin{align}\label{eps}
    \epsilon&= 2 t (\cos p_x + \cos p_y) +2 t^\prime \cos (p_x+p_y).
\end{align}
We can recognize Eq.~\eqref{eps} as the negative of the energy of a single particle hopping on our lattice.  When $t^\prime<t/2$ the ground state occurs at the unique wave-vector ${\vect{p}}=(\pi,\pi)$, corresponding to a uniform spin texture.

When $t^\prime>t/2$ there is a bifurcation, and the energy has minima at ${\vect{p}}=(\pi\pm q/2,\pi\pm q/2)$, with $q=2\cos^{-1} t/(2t^\prime)$.  In that regime the most general texture is {\color{black} a linear superposition of the four degenerate solutions,}
\begin{align}\label{uv}
    \tilde u_j &= e^{i {\vect{p}\cdot \vect{r}_j}} u_+ + e^{-i {\vect{p}\cdot \vect{r}_j}} u_-,\\\nonumber
    \tilde v_j &= e^{i {\vect{p}\cdot \vect{r}_j}} v_+ + e^{-i {\vect{p}\cdot \vect{r}_j}} v_-,
\end{align}
with $|u_+|^2+|u_-|^2+|v_+|^2+|v_-|^2=1/N_s$, where $N_s$ is the total number of sites.  

A convenient representation of these textures comes from using invariance under the following operations (where $\xi=e^{i\phi}$ is a complex number of unit magnitude): 
\begin{enumerate}[(I)]
    \item Gauge transformations: \newline $(u_+,v_+,u_-,v_-)\to \xi(u_+,v_+,u_-,v_-)$
    \item Rotations about the $z$--axis: \newline $(u_+,v_+,u_-,v_-)\to (\xi u_+,\xi^* v_+,\xi u_-,\xi^* v_-)$,
    \item Rotations about the $ x$--axis: \newline $\left(\begin{array}{c}u_\pm\\v_\pm\end{array}\right) \to \left(\begin{array}{cc}
    \cos\theta/2&\sin\theta/2\\
    -\sin\theta/2&\cos\theta/2\end{array}\right)\left(\begin{array}{c}u_\pm\\v_\pm\end{array}\right).$
\end{enumerate}
By combining these operations we can transform any texture of the form of Eqs.~(\ref{uv})  into one with 
$u_+=\cos(\alpha/2) {\color{black} e^{i\gamma}}, u_-=\sin(\alpha/2)\cos(\beta), v_-= -\sin(\alpha/2)\sin(\beta)$.  An $x$--$y$ spiral corresponds to $\alpha=\pi/2,\beta=\pi/2$.  In the main text we work with a $x$--$z$ spiral, which is a rotation of this state.  This ansatz also describes canted spirals, with $\beta=\pi/2$ and $\alpha\neq \pi/2$, and modulated ferromagnets, with $\beta=0$. {\color{black} In all cases $\gamma$ corresponds to a spatial translation of the pattern.
If $q$ is incommensurate with the lattice (as is relevant for the $q\to 0$ limit) then we also have translational invariance,  and the energy will be independent of $\gamma$.   We therefore take $\gamma=0$ for the remainder of this section.}

To analyze the fluctuations about these textures, we rotate our local basis, defining
$a_j = (-\tilde v_j^* c_{j\uparrow} +
\tilde u_j^* c_{j\downarrow})/w_j$ and
$b_j = (\tilde u_j c_{j\uparrow}+ 
\tilde v_j c_{j\downarrow})/w_j$. The factor in the denominator, 
\begin{align}
    w_j=\sqrt{|\tilde u_j|^2+|\tilde v_j|^2},
\end{align}
is necessary for the transformation to be unitary.
The mean-field state in Eq.~(\ref{psi0})
becomes
$
|\psi\rangle= \sum_i w_i  b_i |\Phi\rangle,$
where $|\Phi\rangle=\prod_j  b_j^\dagger |\rm vac\rangle${\color{black}, where $|w_i|^2$ is proportional to the probability of the hole being on site $i$.}  The Hamiltonian then takes the form ${\cal H}={\cal H}_0+{\cal H}^\prime$ where 
\begin{align}\label{H0}
    {\cal H}_0&=-\sum_{ij} t_{ij} \frac{\tilde u_i^* \tilde u_j+\tilde v_i^* \tilde v_j}{w_i w_j} ( a_i^\dagger  a_j +  b_i^\dagger  b_j) \\
    &\equiv -\sum_{ij} \tau_{ij} ( a_i^\dagger  a_j +  b_i^\dagger  b_j)  \nonumber, \\
    \label{Hp}
    {\cal H}^\prime &= -\sum_{ij} t_{ij} \frac{\tilde v_i \tilde u_j-\tilde u_i \tilde v_j}{w_i w_j}  b_i^\dagger  a_j + {\rm h.c.} \\
    &\equiv -\sum_{ij} \lambda_{ij}  b_i^\dagger  a_j + {\rm h.c.}, \nonumber
\end{align}
along with a hard-core constraint.  These reduce to Eqs.~\eqref{h0} and \eqref{hp} in the special case of an  $x$--$z$ spiral.
By construction $|\psi\rangle$ is an eigenstate of ${\cal H}_0$ with eigenvalue $\epsilon$, and 
$\langle \psi|{\cal H}^\prime|\psi\rangle=0$.  
{\color{black}
In the long-wavelength limit, the coefficients become 
\begin{align}
\lambda_{ij} &=i({\vect{q}\cdot \vect{r}}_{ij} )t_{ij}\frac{\sin(\alpha)\sin(\beta) }{1+\sin(\alpha)\cos(\beta) \cos {(\vect{q}\cdot \vect{R}}_{ij})}+{\cal O}(q^2)\nonumber\\
\tau_{ij}&=t_{ij} +{\cal O}(q)\label{expanded}
\end{align}
where ${\vect{ r}}_{ij}={\vect{r}}_i-{\vect{r}}_j$ and ${\vect{R}}_{ij}={\vect{ r}}_i+{\vect{r}}_j$.  One cannot expand the $\cos(\vect{ q}\cdot \vect{R}_{ij})$ term in the denominator of $\lambda_{ij}$, as ${\vect{ R}}_{ij}$ can become arbitrarily large.  Nonetheless, ${\cal H}^\prime$ vanishes as $q\to 0$, and we can use second-order perturbation theory to express the energy as $E=\epsilon+\Delta E$, where
%
\begin{align}
\epsilon&=-2t(\cos q_x + \cos q_y) + 2t^\prime \cos(q_x+q_y),\nonumber\\
    \Delta E&= \langle \psi|{\cal H}^\prime(\epsilon_0-{\cal H}_0)^{-1} {\cal H}^\prime|\psi\rangle,
    \label{de}
\end{align}
and $\epsilon_0=\epsilon(q=0)=-4t+2t^\prime$.  To calculate $\Delta E$ to quadratic order in $q$ one only needs to use the approximate matrix elements from Eq. (\ref{expanded}), which give ${\cal H}_0$ to zeroth order in $q$ and ${\cal H}^\prime$ to linear order.
To that order,
\begin{align}
{\cal H}^\prime |\psi\rangle&=\sum_k {\cal W}_k |\chi_k\rangle,
\end{align}
where $|\chi_k\rangle$ is a (non-normalized) state containing a hole and spin-flip, carrying total momentum $\vect{ k}$,
\begin{align}
|\chi_k\rangle &=\frac{1}{\sqrt{N_s}} \sum_{ij} t_{ij} ({\vect{ q}\cdot \vect{r}}_{ij}) e^{-i{\vect{ k}\cdot \vect{r}}_j} a_i^\dagger b_i b_j |\Phi\rangle.
\end{align}
The coefficient ${\cal W}_k$ encodes the amplitude of that state in the perturbed wavefunction.  It is given by the Fourier transform of the real-space function
\begin{align}
W_j&=-i\frac{\sin(\alpha)\sin(\beta)}{\sqrt{1+\sin(\alpha)\cos(\beta) \cos  {(2\vect{ q}\cdot \vect{r}}_j)}}\\
&=\frac{1}{\sqrt{N_s}}\sum_k
e^{i{\vect {k}\cdot \vect{r}}_j} {\cal W}_k.
\end{align}
 All of the $\alpha$ and $\beta$ dependence  of $\Delta E$ is contained in ${\cal W}_k$.

By translational invariance we then have that $\langle \chi_{k^\prime}|{\cal H}_0 |\chi_k\rangle=0$ unless $k=k^\prime$, and hence
\begin{align}
\Delta E &= \frac{1}{N_s}\sum_k |{\cal W}_k|^2 R_k,
\\
R_k&=\langle \chi_k|(\epsilon_0-{\cal H}_0)^{-1}|\chi_k\rangle.
\end{align}
As $q\to0$,  ${\cal W}_k$ becomes peaked about small $k$. Thus, one can replace $R_k$ with $R_0$ in that sum, implying that $\Delta E\propto \sum_k |{\cal W}_k|^2$.  By Parseval's theorem,
\begin{align}
\Delta E &\propto -\frac{1}{N_s}\sum_j |W_j|^2
\propto -\frac{\sin^2(\alpha)\sin^2(\beta)}{
\sqrt{1-\sin^2(\alpha)\cos^2(\beta)}
},\label{dew}
\end{align}
whose amplitude has a maximum at $\alpha=\beta=\pi/2$. This implies that the spin spiral is the lowest-energy texture.

It remains to be shown that, for a fixed $|q|$, the matrix element $R_0$ is maximized when $q_x=q_y$ and not along some other direction. We demonstrate this in Section~\ref{sec:Variational} by calculating and comparing the spin stiffnesses for $q_x=q_y$ and $q_x=-q_y$.}

\begin{figure}[t!]
    \centering        
    \includegraphics[width=\columnwidth]{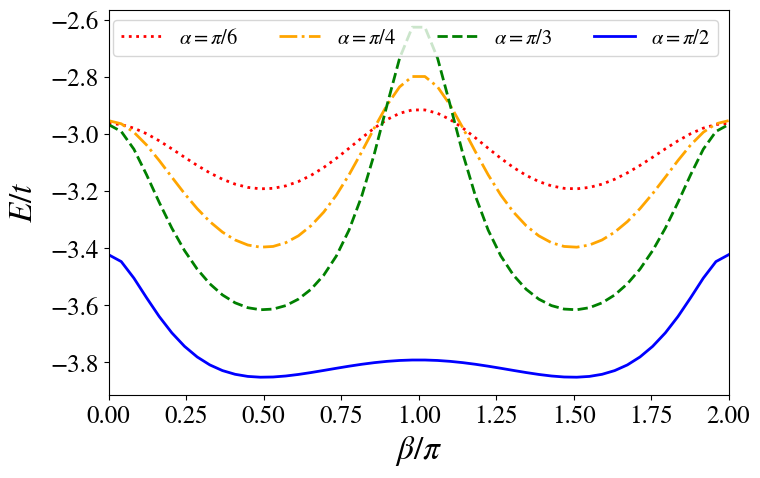}
    \caption{Variational energies of generic spin textures, parameterized by the angles $\alpha$ and $\beta$, {\color{black} as defined in Eq. (\ref{uv}) and the following discussion.  We consider the triangular lattice limit $t^\prime=t$ and take $Q=(2\pi/3,2\pi/3)$.}  The energies are calculated by solving Eqs.~\eqref{eq:NUHElements1} and \eqref{eq:NUHElements2} on a small $15 \times 15$ spatial grid, as a function of $\beta$ and for different choices of $\alpha$: $\pi/6$ (red dotted line), $\pi/4$ (orange dash-dotted line), $\pi/3$ (green dashed line), and $\pi/2$ (blue solid line). The minimum-energy configuration is for $\alpha=\pi/2$ and $\beta=\pi/2$, corresponding to a spin spiral.}
    \label{fig:NonUniform}
\end{figure}

 {\color{black}
\subsection{Optimal Texture at $q=2\pi/3$}\label{global}









The analytic arguments in Section~\ref{quad} only apply in the limit $q\to0$. Here we numerically confirm that for the triangular lattice, $t^\prime=t$, the spin spiral with $q=2\pi/3$ is a global energy minimum within our variational space. This suggests that the conclusions from our earlier analytic arguments persist even when $q$ is large.

Similar to the main text, we make the ansatz}
\begin{align}
    |\psi_2\rangle&=
    \sum_j F_j |j\rangle+
    \sum_{ij} f_{ij} |ij\rangle,
\end{align}
where  $|i\rangle =  b_i |\Phi\rangle$ and $|ij\rangle = a_j^\dagger b_j b_i |\Phi\rangle$ correspond to the state with a hole at location $i$ and a spin flip at $j$.   {\color{black} The spin texture is parameterized by $\alpha,\beta$ as defined in Section~\ref{quad}.}  
Minimizing $\langle \psi_2 | \ham|\psi_2\rangle$ with the constraint $\langle \psi_2|\psi_2\rangle=1$ yields the equations
\begin{align}
    E F_j&= \sum_{k} \tau_{jk}^* F_k -\lambda_{kj} f_{kj},  \label{eq:NUHElements1}\\
    Ef_{ij} &= \sum_{k} (\lambda_{ji}^* F_j + f_{ji} \tau_{ji}) \delta_{kj}+  f_{kj} (1-\delta_{ij}) \tau_{ik}^*. \label{eq:NUHElements2}
\end{align}   
We discretize space and solve these on a finite $15\times 15$ grid, taking $q = 2\pi/3$, $t^\prime = t$, and $\vect{p}=(2\pi/3,2\pi/3)$.  The parameters $\tau_{ij}$ and $\lambda_{ij}$ are defined in Eq.~(\ref{H0}) and (\ref{Hp}). Figure~\ref{fig:NonUniform} shows the resulting energy as a function of $\beta$ for several different values of $\alpha$.  As anticipated, the minimum is at $\beta=\pi/2,\alpha=\pi/2${\color{black}, though the landscape deviates significantly from the long-wavelength prediction in Eq.~(\ref{dew}).  Most notably, there is an asymmetry between $\beta=0$ and $\beta=\pi$, which is related to commensurability of the textures.  {\color{black} We have repeated the calculation with  other parameters, and verified that in the long-wavelength limit the results approach Eq.~(\ref{dew}).}}

\section{Calculating the Energy Integrals \& Spin Stiffnesses}
\label{sec:Variational}

In this section, we explain how to evaluate the energy integrals in Eq.~\eqref{int}, and use them to solve Eq.~(\ref{fs}) for the variational coefficients $f_{\vect{s}}$.  We then determine the corresponding spin stiffness. 

{\color{black} At long wavelengths the energy can be expressed as a quadratic form in $q_x$ and $q_y$.  By symmetry the principle axes are in the direction $q_x=q_y$ and $q_x=-q_y$.  We separately consider these two directions in  Section~\ref{parastiffness}
and Section~\ref{orthostiffness}.  We conclude that the the dominant instability occurs with $q_x=q_y$.}


\subsection{Spin Stiffness along $\mathbf{\textit{Q}} = (q,q)$} \label{parastiffness}
Using the symmetries discussed in the main text, Eq.~(\ref{fs}) can be simplified to a set of two coupled equations,
\begin{align}
    q\left(\begin{array}{c}
    1/2 \\ 1
    \end{array}
    \right)=
    \left(\begin{array}{cc}
    { 1}-ta & t^\prime b \\
    -2tb & { 1}+t^\prime c
    \end{array}
    \right)
    \left(\begin{array}{c}
    f_{\vect{x}} + q/2 \\f_{\vect{w}} + q
    \end{array}
    \right), \label{eq:ParaEqs}
\end{align}
where
\begin{align}
\label{aeq}
    a &= \Lambda_{(0,0)} - \Lambda_{(2,0)} +\Lambda_{(1,-1)} - \Lambda_{(1,1)}, \\
    b &= \Lambda_{(1,0)} - \Lambda_{(2,1)}, \\
    c &=  \Lambda_{(0,0)} - \Lambda_{(2,2)},
\end{align}
and the integrals are defined by Eq.~(\ref{int}).
These equations can be solved as  $f_{\vect{x}}=X q$ and $f_{\vect{w}}=W q$ with
\begin{align}
    X&= -\frac{1}{2} + \frac{ 1}{2}\frac{{1}+t^\prime(c-2b)}{({ 1}-ta)({ 1}+t^\prime c)+2 t t^\prime b^2},\\
    W &= -1+\frac{{ 1}-t(a-b)}{({ 1}-ta)({1}+t^\prime c)+2 t t^\prime b^2}.
\end{align}
Substituting these expressions into the relationship $E=\epsilon_0+q^2(t/2-t^\prime)-\sum_{\vect{u}} S_{\vect{u}} f_{\vect{u}}=\epsilon_0 + q^2 E_2(t,t^\prime)$
gives
\begin{align}
    E_2 = \left(\frac{1+4 X}{2}\right)t-(1+2W) t^\prime.
\end{align}

To evaluate the required integrals, we make  the change of variables $z=e^{i p_x}$ to express
\begin{align}
    \Lambda_{\vect{s}} 
    %
    &= -\int \frac{dp_y}{2\pi} (e^{ip_y})^{s_y} \oint_C \frac{dz}{2\pi i} \frac{z^{s_x}}{z(\epsilon_{\vect{p}}-\epsilon_0)},
\end{align}
where
the contour $C$ is the unit circle
and $\epsilon_{\vect{p}}= -2t \cos p_y
-t(z+z^{-1})-t^\prime (z e^{i p_y}+z^{-1} e^{-i p_y})
$. Given that  we only need to consider $s_x \geq 0$ to calculate $a,b,$ and $c$, the only poles for the contour integral come from the denominator. We write
$z\epsilon_{\vect{p}}-zE = A(z-z_+)(z-z_-)$ with $A=-t+t^\prime e^{i p_y}$ and $z_{\pm} =(-B\pm\sqrt{B^2-A A^*})/A$, where $B=-t\cos p_y-\epsilon_0/2$.  One sees that $z_+ z_-=A^*/A$ has unit norm, and hence only one pole ($z_+$) falls within the contour, yielding a one-dimensional integral which can easily be calculated numerically:
\begin{align}
    \Lambda_{\vect{s}} = \Lambda_{(s_x,s_y)} &= -\int_0^{2\pi} \frac{dp_y}{2\pi} \frac{z_+^{s_x}}{A(z_+-z_-)} (e^{ip_y})^{s_y}.
\end{align}

{\color{black}
\subsection{Spin Stiffness along $\bar{\mathbf{\textit{Q}}} = (q,-q)$} \label{orthostiffness}

The calculation of the spin stiffness along $\bar{\vect{Q}} = (q,-q)$ follows in much the same way. We will mirror the notation of Section~\ref{parastiffness}, placing bars over the various quantities (e.g. $E_2\to\bar E_2$) to distinguish them from the case where  ${\vect{Q}} = (q,q)$.
Equations~(\ref{eq:Helements1}) through (\ref{int}) still hold;
the only change is the wave-vector at which the coefficients $S_{\vect{u}}$ are evaluated.  In this case $\bar f_w=\bar f_{-w}=0,$ and  $\bar f_x=-\bar f_y=-\bar f_{-x}=\bar f_{-y}$.  We find 
$\bar{f}_{\vect{x}} = \bar{X}q$, where
\begin{equation}
    \bar{X} = -\frac{1}{2} + \frac{1}{2}\frac{1}{(1-t\Bar{a})}
\end{equation}
and
\begin{align}
    \bar{a} &= \Lambda_{(0,0)} - \Lambda_{(2,0)} -(\Lambda_{(1,-1)} - \Lambda_{(1,1)}).
\end{align}
%
Note the minus signs compared to Eq.~(\ref{aeq}).
The integrals $\Lambda_{\vect{s}}$ are identical to those previously performed, but are now combined with different coefficients.

For deformations along the $(q,-q)$ direction, the energy is $\bar{E}=\epsilon_0+tq^2/2-\sum_{\vect{u}} S_{\vect{u}} \bar{f}_{\vect{u}}=\epsilon_0 + q^2 \bar{E}_2(t,t^\prime)$. The spin stiffness is therefore
\begin{equation}
    \bar{E}_2 = \left(\frac{1+4 \bar{X}}{2}\right)t.
\end{equation}
We numerically calculate $\bar{E}_2$ and compare it to $E_2$, finding that $\bar{E}_2 > E_2$ for  all $t^\prime \in [0,t/2]$. This proves that spin deformations along the $(q,q)$ direction are energetically preferable over the $(q,-q)$ direction.

It is instructive to re-write both spin stiffnesses,
\begin{align}
    E_2 &= \left(\frac{t}{2}-t^\prime\right) + \left(2Xt-2Wt^\prime\right), &(q_x&=q_y)\\
    \bar{E}_2 &= \left(\frac{t}{2}\right) + (2\bar{X}t). &(q_x&=-q_y)
\end{align}
In each case, the first terms in parentheses capture the dispersion contributions to $E_2$ and $\bar{E}_2$ in the limit of small $q$; the second terms  (involving $X$, $W$, and $\bar{X}$) capture the fluctuation contributions. We find that the fluctuation contributions to $E_2$ and $\bar{E}_2$ are similar in magnitude to one another,
meaning that spin fluctuations are roughly isotropic. Hence, it is the dispersion contribution $(t/2-t^\prime)$ versus $(t/2)$ which favors the $(q,q)$ spin-spiral orientation over the $(q,-q)$ texture.}
\end{document}